\documentclass[12pt]{article}

\def\comp{{\rm C}\llap{\vrule height7.1pt width1pt depth-.4pt\phantom t}}
\def\square{\kern1pt\vbox{\hrule height 1.2pt\hbox{\vrule width 1.2pt\hskip 3pt
   \vbox{\vskip 6pt}\hskip 3pt\vrule width 0.6pt}\hrule height 0.6pt}\kern1pt}
\def\gtwid{\mathrel{\raise.3ex\hbox{$>$\kern-.75em\lower1ex\hbox{$\sim$}}}}
\def\ltwid{\mathrel{\raise.3ex\hbox{$<$\kern-.75em\lower1ex\hbox{$\sim$}}}}
\def\square{\kern1pt\vbox{\hrule height 1.2pt\hbox{\vrule width 1.2pt\hskip 3pt
   \vbox{\vskip 6pt}\hskip 3pt\vrule width 0.6pt}\hrule height 0.6pt}\kern1pt}

\def \be{\begin{equation}}
\def \ee{\end{equation}}
\def \bea{\begin{eqnarray}}
\def \eea{\end{eqnarray}}

\def \del{\partial}

\def \f{\frac}

\def \d{\delta}
\def \D{\Delta}
\def \e{\eta}

\begin{document}

\begin{titlepage}

\begin{flushright}
UFIFT-QG-14-06
\end{flushright}

\vskip 2cm

\begin{center}
{\bf Excitation of Photons by Inflationary Gravitons}
\end{center}

\vskip 2cm

\begin{center}
C. L. Wang$^{*}$ and R. P. Woodard$^{\dagger}$
\end{center}

\begin{center}
\it{Department of Physics, University of Florida, Gainesville, FL 32611}
\end{center}

\vspace{1cm}

\begin{center}
ABSTRACT
\end{center}
We use a recent result for the graviton contribution to the one loop
vacuum polarization to solve the effective field equations for dynamical
photons on de Sitter background. Our results show that the electric field
experiences a secular enhancement proportional to the number of 
inflationary e-foldings. We discuss the minimum this establishes for
primordial inflation to seed cosmic magnetic fields.

\begin{flushleft}
\end{flushleft}

\vskip 2cm

\begin{flushleft}
$^{*}$ e-mail: clwang@ufl.edu \\
$^{\dagger}$ e-mail: woodard@phys.ufl.edu
\end{flushleft}

\end{titlepage}

\section{Introduction}\label{intro}

Because photons have zero mass it does not take much to
affect the long wavelength modes. This has lead many to 
suspect that the explosive expansion of spacetime during
primordial inflation might help to explain cosmic magnetic 
fields \cite{Dolgov:2001ce}. However, the Maxwell Lagrangian 
is conformally invariant, which means that free photons 
cannot locally sense the expansion of spacetime. The search
for an inflationary connection has prompted investigations 
of explicit conformal breaking terms which might be present 
in the effective action \cite{Turner:1987bw}. Quantum 
effects from the conformal anomaly have been also studied 
\cite{Dolgov:1981nw}.

Conformal breaking from other particles can be communicated 
to photons. No one knows the gravitational couplings of
the charged partners of the Standard Model Higgs (which 
become the longitudinal polarizations of the $W^{\pm}$ at 
low energies) but it has been suggested that the 
inflationary production of minimally 
coupled Higgs scalars could endow the photon with a mass 
during inflation \cite{Davis:2000zp,Tornkvist:2000js,
Dimopoulos:2001wx}, and that this might seed the ubiquous 
cosmic magnetic fields of the current epoch 
\cite{Kronberg:1993vk,Grasso:2000wj}. An explicit one loop
computation of the massless charged scalar contribution to 
the vacuum polarization on de Sitter background 
\cite{Prokopec:2002jn,Prokopec:2002uw} has confirmed the 
photon mass conjecture \cite{Prokopec:2003iu}, although 
more work needs to be done to connect this to cosmic 
magnetic fields \cite{Prokopec:2003bx}. Similar one loop 
results pertain as well when the scalar has a small mass
\cite{Prokopec:2003tm,Prokopec:2004au}. 

Because inflation produces more and more charged scalars as 
time progresses (provided they are light and nearly minimally
coupled) the effective photon mass grows. The scalar mass 
remains small during this process \cite{Kahya:2005kj,
Kahya:2006ui,Prokopec:2006ue} until a static, nonperturbative 
limit is eventually reached \cite{Prokopec:2007ak}. The 
vacuum energy drops while this occurs \cite{Prokopec:2008gw} 
and there are dramatic changes in the electrodynamic forces
exerted by point charges and current dipoles 
\cite{Degueldre:2013hba}.

The effects of charged, minimally coupled scalars are 
fascinating but dependent upon assumptions about the unknown
conformal coupling of the Higgs. Gravitons also break conformal
invariance so they too can communicate the violence of primordial 
inflation to the photon sector \cite{Lifshitz:1945du,
Grishchuk:1974ny,Starobinsky:1979ty}. Graviton effects are weaker 
because they are mediated through derivative interactions, but 
they are universal. Hence they serve to establish the minimum 
level at which primordial inflation {\it must} affect 
electromagnetism. The purpose of this paper is to complete the 
derivation of these minimum effects.

Our technique is based on a recent dimensionally regulated and 
fully renormalized computation of the one loop graviton 
contribution to the vacuum polarization $i[\mbox{}^{\mu}
\Pi^{\nu}](x;x')$ on de Sitter background \cite{Leonard:2013xsa}. 
We use this to quantum correct Maxwell's equation,
\begin{equation}
\partial_{\nu} \Bigl[ \sqrt{-g} \, g^{\nu\rho} g^{\mu\sigma}
F_{\rho\sigma}(x)\Bigr] + \int \!\! d^4x' \Bigl[\mbox{}^{\mu}
\Pi^{\nu}\Bigr](x;x') A_{\nu}(x') = J^{\nu}(x) \; , \label{QMax}
\end{equation}
where $g_{\mu\nu}$ is the de Sitter metric, $F_{\rho\sigma} 
\equiv \partial_{\rho} A_{\sigma} - \partial_{\sigma} A_{\rho}$ 
is the usual field strength tensor and $J^{\mu}(x)$ is the
current density. With $J^{\mu}(x) \neq 0$ one can study how 
inflationary gravitons alter the electrodynamic response to 
standard sources, as has recently been done for a point charge
and for a point magnetic dipole with the following results
\cite{Glavan:2013jca}:
\begin{itemize}
\item{An observer co-moving with respect to the sources (hence at 
an exponentially increasing physical distance) perceives the 
magnitude of the point charge to increase linearly with co-moving 
time and logarithmically with the co-moving position;}
\item{The co-moving observer reports only a negative logarithmic
spatial variation in the one loop field of the magnetic dipole;}
\item{An observer at fixed invariant distance from the sources 
perceives no secular change of the point charge; and}
\item{The static observer reports a secular enhancement of the 
magnetic dipole moment.}
\end{itemize}
For our study we set $J^{\mu}(x) = 0$ and work out the one loop
corrections to dynamical photons.

This paper has four sections of which the first is this 
Introduction. In section \ref{effeqn} we use the vacuum 
polarization \cite{Leonard:2013xsa,Glavan:2013jca} to derive an 
equation for the one loop correction to spatial plane wave photon 
mode functions. This equation is solved in section \ref{solve}. 
In section \ref{discuss} we discuss the minimum our result sets
for inflation to seed cosmic magnetic fields.

\section{Effective Mode Equation for Photons}\label{effeqn}

The purpose of this section is to convert the quantum corrected
Maxwell equation (\ref{QMax}) into a simple relation for the one 
loop corrections to the mode function of a plane wave photon on
de Sitter background. We first specialize to plane wave photons
at one loop order. Then we discuss the restrictions imposed by 
cosmology, by effective field theory and by our lack of knowledge 
about the initial state.

\subsection{Perturbative Formulation}\label{pert}

We work on spatially flat sections of de Sitter in conformal 
coordinates,
\begin{equation}
d s^2 = a^2(\eta) 
\left( -d \eta^2 + d\vec{x} \cdot d\vec{x} \right) \, ,
\end{equation}
where $a(\eta)=-\frac{1}{H\eta}=e^{Ht}$ is the scale 
factor and $H$ is the Hubble parameter. Hence the metric
can be written as $g_{\mu\nu} = a^2 \eta_{\mu\nu}$, where 
$\eta_{\mu\nu}$ is the Minkowski metric. Because the Maxwell
Lagrangian is conformally invariant, all the scale factors
cancel in the leftmost term of (\ref{QMax}) and we can 
express it as $\partial_{\nu} F^{\nu\mu}(x)$, where we
raise and lower indices with the Minkowski metric, 
$F^{\nu\mu} \equiv \eta^{\nu\rho} \eta^{\mu\sigma}
F_{\rho\sigma}$.  

We follow \cite{Leonard:2013xsa} in employing a noncovariant 
representation for the vacuum polarization \cite{Leonard:2012si},
\begin{eqnarray}
\lefteqn{ 
i\left[ {}^{\mu} \Pi^{\nu}\right](x;x') = 
\left( \eta^{\mu\nu} \eta^{\rho\sigma} \!-\! 
\eta^{\mu\sigma} \eta^{\nu\rho} \right) 
\partial_{\rho} \partial_{\sigma}' F(x;x')  
}
\nonumber \\
& & \hspace{5cm} +\left( \overline{\eta}^{\mu\nu} 
\overline{\eta}^{\rho\sigma} \!-\! \overline{\eta}^{\mu\sigma} 
\overline{\eta}^{\nu\rho} \right) \partial_{\rho} 
\partial_{\sigma}' G(x;x') \; , \qquad \label{F+G}
\end{eqnarray}
where $\overline{\eta}^{\mu\nu} \equiv 
\eta^{\mu\nu} + \delta^{\mu}_0 \delta^{\nu}_0$ is the purely 
spatial part of the Minkowski metric. (For the transformation
to a covariant representation see \cite{Leonard:2012ex}.)
Substituting (\ref{F+G}) into (\ref{QMax}) with $J^{\mu}=0$
and partially integrating the primed derivatives on the right 
hand side gives, 
\begin{equation}
\partial_{\nu} F^{\nu\mu}(x) = 
-\partial_{\nu} \! \int \!\! d^4x' 
\Biggl\{ i F(x;x') F^{\nu\mu}(x') + 
i G(x;x') F^{\overline{\nu} \overline{\mu}}(x') \Biggr\}
\, . \label{QMax3}
\end{equation}
Here a barred index on any tensor means that its 0-components 
vanish, for example, $V^{\overline{\mu}} \equiv
\overline{\eta}^{\mu\nu} V_{\nu} = V^{\mu} - \delta^{\mu}_0 V^0$.

Some nonperturbative statements can be made. For example, so long
as the vacuum polarization is computed (as the one loop correction
was \cite{Leonard:2013xsa}) using electromagnetic and gravitational
gauge conditions which respect homogeneity and isotropy then the
structure functions $F(x;x')$ and $G(x;x')$ can depend upon the
spatial coordinates $\vec{x}$ and $\vec{x}'$ only through the
Euclidean norm of their difference $\Delta \vec{x} \equiv \vec{x}
- \vec{x}'$. If spatial surface terms vanish (which will be shown
in the next sub-section) we can reflect external space derivatives
onto the field strengths inside the integral of expression 
(\ref{QMax3}),
\begin{eqnarray}
\partial_{\nu} F^{\nu\mu}(x) & = & -\partial_0 \! \int \!\! d^4x'
\, i F(x;x') F^{0\mu} \nonumber \\
& & \hspace{1cm} - \int \!\! d^4x' \Biggl\{ i F(x;x') 
\partial_j' F^{j\mu}(x') + i G(x;x') \partial_j' 
F^{j \overline{\mu}}(x') \Biggr\} \, . \qquad \label{QMax4}
\end{eqnarray}

One simple consequence of (\ref{QMax4}) is the validity, to all 
orders, of the general form for the classical field strengths of 
plane wave photon solutions with wave vector $\vec{k}$ and 
transverse polarization vector $\varepsilon^i(\vec{k},\lambda)$,
\begin{equation}
F^{0i}_{\rm ph}(x) = -\partial_0 u(\eta,k) \times \varepsilon^i
e^{i \vec{k} \cdot \vec{x}} \; , \;
F^{ij}_{\rm ph}(x) = u(\eta,k) \times i [k^i \varepsilon^j \!-\!
k^j \varepsilon^i ] e^{i \vec{k} \cdot \vec{x}} \; . 
\label{pwave}
\end{equation}
To see this, first substitute (\ref{pwave}) into the $\mu = 0$
component of equation (\ref{QMax4}) and act the space derivatives
using $\partial_j F^{\mu\nu}_{\rm ph} = i k_j F^{\mu\nu}_{\rm ph}$,
\begin{equation}
i k_j F^{j 0}_{\rm ph}(x) = 0 - \int \!\! d^4x' \, i F(x;x') 
i k_j F^{j0}_{\rm ph}(x') + 0 \; . \label{0comp}
\end{equation}
Exploiting transversality ($k_j \varepsilon^j = 0$) reduces 
expression (\ref{0comp}) to a tautology of the form $0 = 0$. Now
substitute (\ref{pwave}) into the $\mu = i$ component of 
(\ref{QMax4}) and again exploit transversality,
\begin{eqnarray}
\lefteqn{-\Bigl( \partial_0^2 \!+\! k^2\Bigr) u(\eta,k) \times 
\varepsilon^i e^{i \vec{k} \cdot \vec{x}} = \partial_0 \! \int 
\!\! d^4x' \, i F(x;x') \partial_0' u(\eta',k) \times 
\varepsilon^i e^{i \vec{k} \cdot \vec{x}'} } \nonumber \\
& & \hspace{3cm} + \int \!\! d^4x' \, i\Bigl[ F(x;x') + G(x;x')
\Bigr] k^2 u(\eta',k) \times \varepsilon^i e^{i \vec{k} \cdot
\vec{x}'} \; , \qquad \\
& & \hspace{1cm} = \varepsilon^i e^{i \vec{k} \cdot \vec{x}}
\times \Biggl\{ \partial_0 \! \int \!\! d^4x' \, i F(x;x')
\partial_0' u(\eta',k) e^{-i \vec{k} \cdot \Delta \vec{x}}
\nonumber \\
& & \hspace{3.5cm} + k^2 \! \int \!\! d^4x' \Bigl[ i F(x;x') \!+\!
i G(x;x') \Bigr] u(\eta',k) e^{-i \vec{k} \cdot \Delta \vec{x}}
\Biggr\} . \qquad \label{icomp}
\end{eqnarray}
Dispensing with the now-redundant factors of $\varepsilon^i
e^{i \vec{k} \cdot \vec{x}}$ results in effective mode equation 
for $u(\eta,k)$,
\begin{eqnarray}
\lefteqn{\Bigl( \partial_0^2 \!+\! k^2\Bigr) u(\eta,k) = -
\partial_0 \! \int \!\! d^4x' \, i F(x;x') \partial_0' 
u(\eta',k) e^{-i \vec{k} \cdot \Delta \vec{x}} } \nonumber \\
& & \hspace{3.5cm} - k^2 \! \int \!\! d^4x' \Bigl[ i F(x;x') \!+\!
i G(x;x') \Bigr] u(\eta',k) e^{-i \vec{k} \cdot \Delta \vec{x}}
\; . \qquad \label{modeqn}
\end{eqnarray}
 
Because the structure functions can only depend upon the norm of
$\Delta \vec{x}$ it is possible to quite generally reduce the 
right hand side of the effective mode equation (\ref{modeqn}) to
a double integral over $\eta'$ and $r \equiv \Vert \vec{x} \!-\!
\vec{x} \Vert$. However, we must at this point face the fact 
that the structure functions can only be computed to some finite
order in the quantum gravitational loop counting parameter 
$\kappa^2 \equiv 16 \pi G$, 
\begin{eqnarray}
F(x;x') & = & 0 +\kappa^2 F_{(1)}(x;x') + \kappa^4 F_{(2)}(x;x') 
+ \dots \; , \label{Fexpand} \\
G(x;x') & = & 0 +\kappa^2 G_{(1)}(x;x') + \kappa^4 G_{(2)}(x;x') 
+ \dots \; . \label{Gexpand} 
\end{eqnarray}
At this time we have only the one loop results 
\cite{Leonard:2013xsa,Glavan:2013jca}, 
\begin{eqnarray}
\lefteqn{
i F_{(1)} = \f{-1}{8\pi^2}\Bigg\{ \!H^2\left[\ln(a) \!+\! \alpha 
\right] \!-\! \left[\frac{\ln(a)}{3 a^2} \!-\! \frac{\beta}{a^2}\right]
\left(\del^2 \!+\! 2Ha\del_0 \right) \!+\! \f{H}{3a} \del_0 \!\Bigg\}
\d^4(x \!-\! x') } \nonumber \\
& & \hspace{-.5cm} + \frac{a^{-1} \partial^6}{384\pi^3} \Bigg\{
\frac{\theta(\D\e \!-\! \Delta x)}{a'} \left(\ln\!\left[ \f{1}{4} 
H^2 (\D\e^2 \!-\! \Delta x^2)\right] \!-\! 1 \!\right)\!\Bigg\} 
- \f{H^2}{32\pi^3} \Bigg\{\! \left[ \frac{\partial^4}{4} \!+\! 
\partial^2 \partial_0^2 \right] \nonumber \\
& & \hspace{.5cm} \times \theta(\D\e \!-\! \Delta x) 
\ln\left[\f{1}{4} H^2 (\D\e^2 \!-\! \Delta x^2)\right]
- \left[\frac{\partial^4}{4} \!-\! \del^2 \del_0^2 \right]
\theta(\D\e \!-\! \Delta x) \Bigg\} , \qquad \label{finalfr2} \\
\lefteqn{i G_{(1)} = \f{H^2}{6\pi^2} \left[\ln(a) + \frac34 
\gamma \right] \delta^4(x \!-\! x')} \nonumber \\
&&\hspace{2.5cm} + \f{H^2 \partial^4}{96\pi^3} \Bigg\{\theta(\D\e \!-\! 
\Delta x) \left(\ln\left[ \f{1}{4} H^2(\D\e^2 \!-\!
\Delta x^2)\right] - 1 \right)\Bigg\} \; . \qquad \label{finalgr2}
\end{eqnarray}
In these and subsequent expressions the coordinate separations are
$\Delta \eta \equiv \eta - \eta'$ and $\Delta x \equiv \Vert \vec{x}
- \vec{x}'\Vert$ and the flat space d'Alembertian is $\partial^2 \equiv
\eta^{\mu\nu} \partial_{\mu} \partial_{\nu} = -\partial_0^2 + \nabla^2$.

Because the structure functions are only known to a finite order in
$\kappa^2$ there is no alternative to making a similar expansion for
the mode function,
\begin{equation}
u(\eta,k) = u_0(\eta,k) + \kappa^2 u_1(\eta,k) + \kappa^4 u_2(\eta,k)
+ \dots \label{uexpand}
\end{equation}  
Substituting expansions (\ref{Fexpand}-\ref{Gexpand}) and 
(\ref{uexpand}) into (\ref{modeqn}) and segregating terms of the same 
order in $\kappa^2$ gives the tree order and one loop relations,
\begin{eqnarray}
\Bigl( \partial_0^2 \!+\! k^2\Bigr) u_0(\eta,k) & = & 0 \; , 
\label{0kappa} \\
\Bigl( \partial_0^2 \!+\! k^2\Bigr) u_1(\eta,k) & = & 
-\partial_0 \! \int \!\! d^4x' \, i F_{(1)}(x;x') \partial_0' 
u_0(\eta',k) e^{-i \vec{k} \cdot \Delta \vec{x}} \nonumber \\
& & \hspace{-.5cm} - k^2 \! \int \!\! d^4x' \Bigl[ i F_{(1)}(x;x') 
\!+\! i G_{(1)}(x;x') \Bigr] u_0(\eta',k) e^{-i \vec{k} \cdot \Delta 
\vec{x}} \; . \qquad \label{SQmax}
\end{eqnarray}
By conformal invariance the tree order mode function is the same in
de Sitter conformal coordinates as it is in flat space, $u_0(\eta,k) 
= e^{-ik \eta}/\sqrt{2 k}$. 

\subsection{Schwinger-Keldysh Formalism}\label{SK}

The treatment (though not of course the explicit structure functions
(\ref{finalfr2}-\ref{finalgr2})) we have given so far applies as well 
for traditional quantum field theory on flat space (for example, see 
\cite{Leonard:2012fs}). However, it is important to understand
that the usual effective field equations describe matrix
elements of the field operator between states which are free in
the asymptotic past and future. These in-out matrix elements 
provide a correct description of scattering processes in flat 
space, but they make little sense in cosmology because the universe 
began with a singularity at some finite time and no one knows how 
(or even if) it will end. Persisting with the in-out effective 
field equations for inflationary cosmology would result in two 
embarrassments from the nonlocal source term on the right hand side 
of expression (\ref{SQmax}):
\begin{itemize}
\item{Because the in-out structure functions do not vanish for
${x'}^{\mu}$ outside the past light-cone of $x^{\mu}$ the right
hand side of \ref{SQmax}) would be dominated by contributions from
the far future when the inflated 3-volume is much larger;}
\item{Because the in-out structure functions are complex they 
couple the real and imaginary parts of the mode function 
$u(\eta,k)$, making real field strengths impossible.}
\end{itemize}

The more meaningful effective field to study for cosmology is the
true expectation value of the field operator in the presence of 
some state which is released at a finite time. The appropriate
field equations for studying expectation values are
those of the Schwinger-Keldysh formalism \cite{Schwinger:1960qe,
Mahanthappa:1962ex,Bakshi:1962dv,Bakshi:1963bn,Keldysh:1964ud,
Chou:1984es,Jordan:1986ug,Calzetta:1986ey,Calzetta:1986cq}. The
associated one loop structure functions were given in expressions
(\ref{finalfr2}-\ref{finalgr2}). Note that they are manifestly
real, and that the factors of $\theta(\Delta \eta - \Delta x)$ make
each term vanish whenever the point ${x'}^{\mu}$ strays outside the
past light-cone of $x^{\mu}$.\footnote{One consequence is that spatial
integration by parts produces no surface terms in the Schwinger-Keldysh
formalism. Partial integration in time can and does produce surface 
terms at the initial time.} These are important features of the
Schwinger-Keldysh formalism which the in-out formalism lacks.

The constants $\alpha$, $\beta$ and $\gamma$ which appear in 
expressions (\ref{finalfr2}-\ref{finalgr2}) represent the arbitrary
finite parts of the three higher derivative counterterms which were
needed to renormalize the vacuum polarization \cite{Leonard:2013xsa}
because Einstein + Maxwell is not perturbatively renormalizable
\cite{Deser:1974zzd,Deser:1974cz}. (Appendix A discusses the 
noncovariant counterterm resulting from the use of a de Sitter breaking
gauge to compute the vacuum polarization \cite{Leonard:2013xsa}.) No 
physical principle can fix these constants because those counterterms 
cannot actually be present in fundamental theory. They are the price we 
must pay for using Einstein + Maxwell as a low energy effective field 
theory. In contrast, the logarithms of the scale factor with which the 
three constants are paired,
\begin{equation}
\ln(a) + \alpha \qquad , \qquad \frac13 \ln(a) - \beta \qquad , 
\qquad \ln(a) + \frac34 \gamma \; , 
\end{equation}
represent unique and reliable predictions of the theory which must
persist in whatever is the correct ultraviolet completion of Einstein
+ Maxwell. At late times these logarithms dwarf the unknown constants,
which means that we can make reliable predictions in the late time
regime. We are of course making the usual assumption of low energy
effective field theory that $\alpha$, $\beta$ and $\gamma$ are of 
order one.

Another limit to the generality of our formalism is that the 
structure functions (\ref{finalfr2}-\ref{finalgr2}) were computed 
without correcting the free vacuum state. For in-out matrix
elements we typically do not worry about correcting the states 
because infinite time evolution is supposed to accomplish this in
the weak operator sense. However, when the universe is released at
a finite time one must include at least the perturbative corrections
to the initial state. In the Schwinger-Keldysh formalism these
corrections show up as new interaction vertices on the initial value
surface \cite{Ford:2004wc}. Unlike the finite parts of the higher
derivative counterterms, it is perfectly possible to work these 
corrections out \cite{Collins:2005nu,Collins:2005cm,Collins:2006bg,
Collins:2006uy,Garny:2009ni,Kahya:2009sz}. However, there is no 
point to doing so because they give rise to surface terms which 
fall off like powers of the inflationary scale factor. We shall 
assume that these corrections simply serve to cancel the surface 
terms which would arise, at various points, from partial 
integrations.

\section{Solving the Equation}\label{solve}

The purpose of this section is to solve equation (\ref{SQmax}) for
$u_1(\eta,k)$ in the late time limit for which reliable predictions 
can be made. First note that (\ref{SQmax}) can be expressed in terms
of seven master integrals,
\begin{eqnarray}
\lefteqn{ (\partial_0^2 \!+\! k^2) u_1(\eta,k) = i k \partial_0 \Bigl[
I_1 \!+\! I_2 \!+\! I_3 \!+\! I_4 \!+\! I_5 \!+\! I_6 \!+\! I_7\Bigr] }
\nonumber \\
& & \hspace{5cm} + k^2 \Bigl[ -I_1 \!+\! \frac13 I_2 \!-\! I_3 \!+\! 
\frac13 I_4 \!-\! I_5 \!-\! I_6 \!-\! I_7\Bigr] \; , \qquad \label{newmax}
\end{eqnarray}
where the various integrals are,
\begin{eqnarray}
I_1(\eta,k) & \!\!\!\!\! \equiv \!\!\!\!\! & -\frac{a^{-1} 
(\partial^2_0 \!+\! k^2)^3}{384\pi^3} \!\! \int \!\! d^4x' 
\frac{\Theta}{a'} \!\left[\ln \!\left[ \frac{H^2}{4} \!\left(\Delta \eta^2 
\!-\! \Delta x^2\right) \!\right] \!-\! 1\right] u_0(\eta',k) e^{-i \vec{k} 
\cdot \Delta \vec{x}} \!\! , \quad \label{integl1} \\
I_2(\eta,k) & \!\!\!\!\! \equiv \!\!\!\!\! & -\frac{H^2 (\partial_0^2
\!+\! k^2)^2}{128\pi^3} \!\! \int\!\! d^4x' \Theta \left[ 
\ln\! \left[\frac{H^2}{4} \!\left(\Delta \eta^2 \!-\! \Delta x^2 \right)
\! \right] \!-\! 1 \right] u_0(\eta',k) e^{-i\vec{k} \cdot \Delta \vec{x}} 
\!\! , \label{integl2} \\
I_3(\eta,k) & \!\!\!\!\! \equiv \!\!\!\!\! & \frac{H^2 (\partial^2_0
\!+\! k^2) \partial_0^2}{32\pi^3} \!\! \int \!\! d^4x' \Theta \!
\left[\! \ln\!\left[\f{H^2}{4} \!\left(\Delta \eta^2 \!-\! \Delta x^2
\right) \! \right] \!+\! 1 \right] u_0(\eta',k) e^{-i\vec{k} \cdot \Delta 
\vec{x}} \!\! , \qquad \label{integl3} \\
I_4(\eta,k) & \!\!\!\!\! \equiv \!\!\!\!\! & -\f{H^2 \ln(a)}{8\pi^2} 
\! \int \!\! d^4x' \delta^4(x \!-\! x') u_0(\eta',k) e^{-i\vec{k} 
\cdot \Delta \vec{x}} , \label{integl4} \\
I_5(\eta,k) & \!\!\!\!\! \equiv \!\!\!\!\! & -\frac{a^{-2} \ln(a)
(\partial^2_0 \!+\! k^2)}{24\pi^2} \! \int \!\! d^4x' \delta^4(x 
\!-\! x') u_0(\eta',k) e^{-i\vec{k} \cdot \Delta \vec{x}} , 
\label{integl5} \\
I_6(\eta,k) & \!\!\!\!\! \equiv \!\!\!\!\! & \f{H a^{-1} \ln(a) 
\partial_0}{12\pi^2} \! \int \!\! d^4x' \delta^4(x \!-\! x') 
u_0(\eta',k) e^{-i \vec{k} \cdot \Delta \vec{x}} , \label{integl6} \\
I_7(\eta,k) & \!\!\!\!\! \equiv \!\!\!\!\! & -\frac{H a^{-1}
\partial_0}{24\pi^2} \! \int \!\! d^4x' \delta^4(x \!-\! x') 
u_0(\eta',k) e^{-i \vec{k} \cdot \Delta \vec{x}} , \label{integl7} 
\end{eqnarray}
To save space, we have defined the causality-enforcing
$\theta$-function as $\Theta \equiv \theta(\Delta \eta - \Delta x)$.
Of course the delta function terms (\ref{integl4}-\ref{integl7}) are 
trivial, and most of the nonlocal contributions can be inferred from 
previous work \cite{Miao:2006gj}. Technical details can be found in
Appendix B but the results are,
\begin{eqnarray}
I_1(\eta,k) & \!\!\!\!\! = \!\!\!\!\! & \frac{H^2 u_0(\eta,k)}{48 \pi^2 a}
\Biggl\{ \Bigl[ \frac{1 \!+\! 2 i k \Delta \eta_i \!+\! e^{2 i k 
\Delta \eta_i}}{H^2 \Delta \eta_i^2}\Bigr] + \Bigl[ \frac{1 \!+\!
e^{2i k \Delta \eta_i}}{H \Delta \eta_i} \Bigr] \nonumber \\
& & \hspace{3.3cm} -\frac{4 i k}{H} \ln(H \Delta \eta_i) - \frac{2 i k}{H}
\!\! \int_0^1 \!\! \frac{dt}{t} \Bigl[ e^{2 i k \Delta \eta_i t} \!-\! 
1\Bigr] \Biggr\} , \qquad \label{integl1ans} \\
I_2(\eta,k) & \!\!\!\!\! = \!\!\!\!\! & \frac{H^2 u_0(\eta,k)}{16 \pi^2}
\Biggl\{ -2 \ln(H \Delta \eta_i) - \int_0^1 \!\! \frac{dt}{t} \Bigl[
e^{2 i k \Delta \eta_i t} \!-\! 1\Bigr] \Biggr\} , \label{integl2ans} \\
I_3(\eta,k) & \!\!\!\!\! = \!\!\!\!\! & \frac{H^2 u_0(\eta,k)}{16 \pi^2} 
\Biggl\{ \Bigl[ 6 \!-\! 4 i k \Delta \eta_i \!+\! 2 e^{2 i k \Delta \eta_i}
\Bigr] \ln(H \Delta \eta_i) \!+\! e^{2 i k \Delta \eta_i} \!+\! 7 \!-\!
2 i k \Delta \eta_i \nonumber \\
& & \hspace{0.3cm} + \int^1_0 \!\! \frac{dt}{t} \Biggl[ (3 \!-\! 
2 i k \Delta \eta_i) (e^{2 i k \Delta \eta_i t} \!-\! 1) \!+\!
e^{2 i k \Delta \eta_i} ( e^{-i 2 i k \Delta \eta_i t} \!-\! 1) \Biggr]
\Biggr\} , \qquad \label{integl3ans} \\
I_4(\eta,k) & \!\!\!\!\! = \!\!\!\!\! & -\f{H^2 \ln(a)}{8\pi^2} \times 
u_0(\eta,k) \; , \label{integl4ans} \\
I_5(\eta,k) & \!\!\!\!\! = \!\!\!\!\! & 0 \; , \label{integl5ans} \\
I_6(\eta,k) & \!\!\!\!\! = \!\!\!\!\! & -\f{i k H \ln(a)}{12\pi^2a} 
\times u_0(\eta,k) \; , \label{integl6ans} \\
I_7(\eta,k) & \!\!\!\!\! = \!\!\!\!\! & \f{i k H}{24\pi^2 a} \times 
u_0(\eta,k) \; . \label{integl7ans}
\end{eqnarray}
Here and henceforth we define $\Delta \eta_i \equiv \eta - \eta_i = 
H^{-1} (1 - \frac1{a})$, where $\eta_i$ is the initial conformal time.

A few comments are in order regarding the inverse factors of $\Delta 
\eta_i$ which appear in expression (\ref{integl1ans}) for $I_1(\eta,k)$.
{\it These factors diverge on the initial value surface and completely
preclude any attempt to exactly solve the one loop truncated effective
field equations in their present form.} That the problem has nothing to 
do with de Sitter background is obvious from the fact that these very 
same divergences appear as well when the background is changed to flat 
space \cite{Leonard:2012fs}. The problem arises instead because the 
vacuum polarization was computed in free (Bunch-Davies) vacuum 
$\Omega_0[A,h]$ \cite{Leonard:2013xsa}. Even in flat space the true 
vacuum state wave functional $\Omega[A,h]$ requires perturbative 
corrections \cite{Collins:2005nu,Collins:2005cm,Collins:2006bg,
Collins:2006uy,Garny:2009ni,Kahya:2009sz},
\begin{eqnarray}
\lefteqn{\Omega[A,h] = \Omega_0[A,h] \times \Biggl\{1 + \kappa \! \int 
\!\! d^3x_1 \!\! \int \!\! d^3x_2 \!\! \int \!\! d^3x_3 } \nonumber \\
& & \hspace{1.5cm} \times 
\Omega^{\mu\nu\rho\sigma}(\vec{x}_1,\vec{x}_2,\vec{x}_1) 
h_{\mu\nu}(\eta_i,\vec{x}_1) A_{\rho}(\eta_i,\vec{x}_2) 
A_{\sigma}(\eta_i,\vec{x}_3) + O(\kappa^2) \Biggr\} , \qquad 
\end{eqnarray}
where $\Omega^{\mu\nu\rho\sigma}(\vec{x}_1,\vec{x}_2,\vec{x}_3)$ is
a $\comp$-number function which could be worked out --- but has not
been --- the same way one computes corrections the simple harmonic 
oscillator wave functions when the Hamiltonian contains an 
anharmonic term. {\it We stress that the problem derives from
combining ($D=4$) interactions with evolution from a finite time,
and it cannot be solved by any clever choice Gaussian initial state
$\Omega_0[A,h]$.} It can be avoided in flat space by taking the
initial time to $-\infty$ \cite{Leonard:2012fs} but this is not an
option for our de Sitter computation owing to the factors of 
$\ln(a)$ and $1/a$ which are evident in expressions 
(\ref{integl4ans}) and (\ref{integl6ans}-\ref{integl7ans}).\footnote{
The physical origin of this mathematical obstacle is inflationary 
particle production which results in very high occupation numbers 
$N(t,k) = [H a(t)/2k]^2$ for graviton modes which have experienced 
first horizon crossing. The earlier one releases the initial state 
the more modes will have experienced first horizon crossing by any 
fixed late time.} On the other hand, the initial value divergences 
in expression (\ref{integl1ans}) are all well behaved at late times,
\begin{equation}
\frac1{H^2 \Delta \eta_i^2} \longrightarrow 1 \;\; , \;\;
\frac1{H \Delta \eta_i} \longrightarrow 1 \;\; , 
\ln(H \Delta \eta_i) \longrightarrow 0 \; .
\end{equation}
The multiplicative factor of $1/a$ makes $I_1(\eta,k)$ go to zero 
at late times, so we can avoid computing initial state corrections
by simply working consistently in the late time regime, which is 
in any case necessary owing to the unknown finite parts of the
counterterms. 

\begin{table}
\centering
\begin{tabular}{|c|c|c|}
\hline 
$i$ & $\sqrt{2k} \, I_i(\eta,k)$ & $\sqrt{2k} \, \partial_0 I_i(\eta,k)$ \\ 
\hline 
1 & $O(\frac1{a})$ & $O(1)$ \\ 
\hline 
2 & $O(1)$ & $O(1)$ \\ 
\hline 
3 & $O(1)$ & $O(1)$ \\ 
\hline 
4 & $-\frac{H^2 \ln(a)}{8\pi^2}$ & $-\f{H^3 a}{8\pi^2}$ \\ 
\hline 
5 & 0 & 0 \\ 
\hline 
6 & $O(\frac{\ln(a)}{a})$ & $\frac{i k H^2 \ln(a)}{12\pi^2}$ \\ 
\hline 
7 & $O(\frac1{a})$ & $O(1)$ \\ 
\hline 
\end{tabular} 
\caption{Leading late time limiting forms for the integrals defined in 
expressions (\ref{integl1}-\ref{integl7}) and their first time derivatives.
Only terms which show secular growth are given explicitly. \label{LTintegl}}
\end{table}

Table \ref{LTintegl} gives the leading late time effect from each of the
seven integrals and its first (conformal) time derivative. The dominant 
effect derives from the time derivative of $I_{4}(\eta,k)$,
\begin{equation}
(\partial_0^2 \!+\! k^2) u_1(\eta,k) = -\frac{i k H^3 a}{8 \pi^2} 
\times u_0(\eta,k) + O\Bigl( \ln(a)\Bigr) \; .
\end{equation}
Hence we find,
\begin{equation}
u_1(\eta,k) = \frac{i k H \ln(a)}{8 \pi^2} \times u_0(\eta,k) +
O\Bigl( \frac1{a}\Bigr) \; . \label{SKSQmax_sol}
\end{equation}
From expression (\ref{pwave}) follows that the one loop field 
strengths are,
\begin{eqnarray}
\kappa^2 F_{(1)}^{0i}(x) & = & \frac{\kappa^2 H^2}{8 \pi^2} 
\Biggl\{\ln(a) + O(1) \Biggr\} \times F_{(0)}^{0i}(x) \; , 
\label{LoopField_0} \\
\kappa^2 F_{(1)}^{ij}(x) & = & \frac{\kappa^2 H^2}{8 \pi^2}
\Biggl\{ \frac{ik \ln(a)}{H a} + O\Bigl(\frac1{a}\Bigr) \Biggr\}
\times F^{ij}_{(0)}(x) \; . \label{LoopField_i}
\end{eqnarray}

\section{Discussion}\label{discuss}

We have employed a previous computation of the one loop 
contribution to the vacuum polarization from inflationary
gravitons \cite{Leonard:2013xsa} to derive what happens to
photons during primordial inflation. Our results 
(\ref{LoopField_0}-\ref{LoopField_i}) for the field strengths
show that the electric field experiences a secular enhancement,
relative to its classical value. In contrast, the one loop
correction to the magnetic field falls off with respect to its 
classical counterpart. Both results are consistent with the one 
loop photon wave function (\ref{SKSQmax_sol}) relaxing to zero 
less slowly (by one factor of $\ln(a)$) than the classical mode 
function approaches a constant. 

The enhancement we find seems to derive from the buffeting of
photons by inflationary gravitons. Even though the photon's
kinetic energy redshifts to zero, its spin does not and this 
permits it to continue interacting with inflationary gravitons
even at late times. The same $\ln(a)$ enhancement was found 
for massless fermions \cite{Miao:2005am,Miao:2006gj,Miao:2007az},
and was explicitly tied to the spin interaction \cite{Miao:2008sp}.
In contrast, massless, minimally coupled scalars neither
experience any significant effect from inflationary gravitons
\cite{Kahya:2007bc,Kahya:2007cm}, nor do they induce a significant
effect on inflationary gravitons \cite{Park:2011ww,Park:2011kg,
Leonard:2014zua}. Gravitons also have spin so it would be very
interesting to see what they do to themselves, as well as to the
force of gravity.

An interesting technical detail concerns the comparison of our
full one loop computations with the result previously derived 
using the Hartree approximation \cite{Leonard:2013xsa}. Both 
calculations give the same time dependence, confirming the 
general reliability of the Hartree approximation for predicting 
the functional form. However, the sign of our exact computation 
differs from that of the Hartree result. This emphasizes the
need for making exact computations, and has clear implications
for gravitons \cite{Mora:2013ypa}.

One important consequence of our result is that quantum 
gravitational perturbation theory must break down after an 
enormous number of e-foldings $\ln(a) \sim 1/\kappa^2 H^2$, 
which is larger than $10^{10}$ in standard inflation. Note first 
that this eventual breakdown in no way invalidates the use of 
perturbation theory at earlier times. The reason for working in 
the late time regime is that we have not included perturbative 
corrections to the initial state (although this could be done) 
and that we do not know, and cannot know, the finite parts 
$\alpha$ and $\beta$ of the higher time derivative counterterms. 
{\it However, as long as one makes the usual assumption that 
these finite parts are of order one it will be seen that there 
is an enormous range of times for which the late time limiting 
form} (\ref{LoopField_0}) {\it vastly dominates the unknown 
contributions, but is still small compared with the classical 
result.} For example, if it is agreed that the infrared logarithm 
$\ln(a)$ dominates when it has reached $\ln(a) \gtwid 100$, and 
that it is still reliable for $\kappa^2 H^2 \ln(a) \ltwid 1/100$, 
then we see that {\it the regime of validity for our result 
extends to a million e-foldings of inflation.} For some 
perspective it should be recalled that there is no currently 
recognized evidence for more than about 60 e-foldings of inflation.

Still, it is a fact that perturbation theory must eventually
break down if inflation persists long enough, and we would
dearly love to know what takes place afterwards. The analogous
problem for scalar potential models on nondynamical de Sitter 
(whose loop corrections also involve factors of $\ln(a)$) can 
be solved using Starobinsky's stochastic formalism 
\cite{Starobinsky:1986fx,Starobinsky:1994bd}. A proof has been 
given that this method reproduces the leading secular growth 
factors at each order in perturbation theory 
\cite{Woodard:2005cw,Tsamis:2005hd}, and it has been extended
to include scalars coupled to fermions \cite{Miao:2006pn} and
to electromagnetism \cite{Prokopec:2007ak}. Detailed analysis
of the nonperturbative resummation of this series of leading
logarithms reveals that all three logical possibilities occur
for the induced vacuum energy:
\begin{itemize}
\item{It can approach a small positive constant 
\cite{Starobinsky:1994bd};}
\item{It can approach a small negative constant 
\cite{Prokopec:2007ak}; or}
\item{It can decrease without bound, resulting in a Big Rip
singularity \cite{Miao:2006pn}.}
\end{itemize}
The stochastic formalism is also very useful in debunking 
techniques which are sometimes proposed for evolving past the 
breakdown of perturbation theory, for example, using order 
reduction to solve the one loop truncated effective field 
equations exactly \cite{Simon:1990ic},\footnote{Order 
reduction is just a technique for preventing higher derivatives 
from introducing new solutions, which problem does not even 
arise in our one loop solution. The problem comes in solving 
the one loop truncated equations exactly. This is only valid if
no equally strong contributions occur in primitive contributions
from higher loops. The stochastic formalism shows that they do.}
or employing variants of the renormalization group \cite{lots}. 
These techniques do give equations which can be evolved past the 
breakdown of perturbation theory, but the results are wrong
\cite{Woodard:2008yt}.

Unfortunately, the presence of derivative interactions means 
that the proof of Starobinsky's formalism which was given for 
scalar potential models \cite{Woodard:2005cw,Tsamis:2005hd} 
does not apply to quantum gravity. One might speculate that 
the stochastic formalism is nonetheless correct, and it has 
been employed on this basis to estimate corrections to the
inflationary power spectra \cite{Urakawa:2008rb}. Other
plausible approximations were earlier invoked to arrive at 
somewhat different estimates \cite{Sloth:2006az,Seery:2007we,
Seery:2007wf}. And there has been recent work by Kitamoto and 
Kitazawa on secular growth corrections to gauge couplings 
\cite{Kitamoto:2012ep,Kitamoto:2012vj,Kitamoto:2013ira,
Kitamoto:2013rea,Kitamoto:2014gva}. The key question is, does 
any specific formalism reproduce the correct secular growth 
factors from inflationary gravitons? Absent a proof, one can 
never know except by comparison with complete and fully 
renormalized results such as the one (\ref{LoopField_0}) we 
have just derived. {\it Indeed, expression} (\ref{LoopField_0}) 
{\it is only the second example of such a result so it 
effectively doubles what is reliably known about this 
fascinating phenomenon.}

An appreciation of the importance of (\ref{LoopField_0}) comes
by recalling what was learned from the only other complete and
fully renormalized graviton loop which shows a factor of 
$\ln(a)$, the 2005-6 computation of inflationary graviton 
corrections to massless fermions \cite{Miao:2005am,Miao:2006gj,
Miao:2007az}. A diagram-by-diagram analysis of that result
reached the following conclusions \cite{Miao:2008sp}:
\begin{itemize}
\item{Naive application of the stochastic formalism does not
correctly predict the secular growth factor of $\ln(a)$;}
\item{The factor of $\ln(a)$ derives entirely from diagrams
which involve the fermion spin connection; and}
\item{Because the ultraviolet sector of the fermion field 
contributes to the factor of $\ln(a)$, one must leave the
ultraviolet regularization on as well for the graviton.}
\end{itemize}   
One should also note the claim that false indications of
secular corrections can come from neglecting diagrams
\cite{Pimentel:2012tw}.

It is not that approximate calculations are necessarily wrong
but rather that we cannot currently judge their validity. When 
a technique is finally devised for isolating the leading secular 
effects at each order and re-summing them it will no doubt lead
to simple approximations for quickly inferring just the leading 
effects, without enduring the tedium of a complete and fully 
renormalized computation. We have reached this point with scalar
potential models through Starobinsky's stochastic formalism.
However, we are not there yet for quantum gravity and we are not
likely to get there, or be sure that we have even made any 
progress, without careful examination of exact results such as
expression (\ref{LoopField_0}). These comparisons sometimes 
reveal subtle problems with plausible-sounding arguments such 
as the universal applicability of the stochastic formalism or 
the irrelevance of ultraviolet effects.

Finally, it is interesting to work out what our result says for 
the possibility of inflation seeding cosmic magnetic fields. We
can use the 0-point energy to estimate the number of photons 
created by inflationary gravitons. Because the co-moving time 
$t$ is related to the conformal time $\eta$ by $dt = a d\eta$, 
the physical Hamiltonian (which generates evolution in $t$) is 
$1/a$ times the conformal one. The physical 0-point energy in
a single photon polarization wave vector $\vec{k}$ is
therefore,
\begin{eqnarray}
\frac1{2a} \Bigl[ F^{0i}_{\rm ph} \times {F^{0i}_{\rm ph}}^* + 
\frac12 F^{ij}_{\rm ph} \times {F^{ij}_{\rm ph}}^* \Bigr] & = &
\frac{k^2 u u^*}{2 a} \; , \\
& \longrightarrow & \frac{k}{2 a} \times \Biggl\{ 1 + 
\frac{\kappa^2 H^2 \ln(a)}{8\pi^2} + O(\kappa^4) \Biggr\} . 
\qquad \label{loopE}
\end{eqnarray}
The occupation number $N(\eta,k)$ is defined by equating
the 0-point energy (\ref{loopE}) to $(N + \frac12) \times 
\frac{k}{a}$,
\begin{equation}
N = \frac{\kappa^2 H^2 \ln(a)}{16 \pi^2}
+ O(\kappa^4) \; . \label{Num}
\end{equation}
The remainder of the computation is the same as the analysis
\cite{Prokopec:2003bx} for the much larger effect from 
inflationary scalars (if the Higgs is minimally coupled and 
still light at inflationary energy scales). Substituting our
value (\ref{Num}) for the occupation number into equation
(118) of that paper results in the following estimate for the 
magnetic strength on scale $\ell_0$,
\begin{equation}
B^2(\ell_0) \sim \frac{\hbar^2 G H_I^2 N_I}{8 \pi^3 c^4 
\ell_0^4} \; ,
\end{equation}
where $H_I$ is the inflationary Hubble parameter and $N_I$ is
the value reached by $\ln(a)$. Plugging in the numbers with
$N_I \sim 50$ and $\ell_0 \sim 10~{\rm kpc}$ gives,
\begin{equation}
B(\ell_0) \sim 10^{-61}~{\rm Gauss} \; .
\end{equation}
This is far too small to have seeded today's
cosmic magnetic fields, but it does serve to establish the
absolute minimum effect which {\it must} be present from
inflation.

\section{Appendix A: de Sitter Breaking Gauges}\label{break}

Our analysis is based on an earlier computation \cite{Leonard:2013xsa}
of the one loop vacuum polarization from inflationary gravitons that 
was made using electromagnetic and gravitational gauge fixing terms 
which respect homogeneity and isotropy, and also dilatation invariance, 
but not the three remaining symmetries of the de Sitter group. The 
special feature of these gauge fixing terms is that the photon and 
graviton propagators consist of linear combinations of constant tensor 
factors times scalar propagators \cite{Tsamis:1992xa,Woodard:2004ut}.
This makes the computation tractable but it also means that 
renormalization can involve noninvariant counterterms, and one such
does occur \cite{Leonard:2013xsa}.\footnote{The vacuum polarization
was recently computed in a manifestly de Sitter invariant gauge.
Strangely enough, the noninvariant counterterm is still required owing
the the time-ordering of interactions, the fact that gravitational
interactions contain two derivatives, and the fact that the coincident
graviton propagator diverges in de Sitter background
\cite{Glavan:2015ura}.} Similar noninvariant counterterms arose 
from the graviton contribution to the one loop fermion self-energy 
\cite{Miao:2005am} (see equations (74), (216) and (221))and the one 
loop self-mass-squared of of a massless, minimally coupled scalar 
\cite{Kahya:2007bc} (see equation (145) and Table 8).

Those accustomed to modern techniques of renormalization in covariant
gauges sometimes find the appearance of noninvariant counterterms to 
be disconcerting. However, it is important to realize that they pose 
no problem of principle. The divergent part of the counterterm is of 
course fixed by the primitive divergences it is to remove, and the 
finite part can be determined to enforce physical symmetries. (In our 
case the focus on late times obviates the need for this as long as 
the finite part of the counterterm is assumed to be of order one.) 
The procedure is explained in older standard texts on quantum field 
theory, for example \cite{Jauch:1955}. And it is important to 
recognize that many of the classic computations of quantum
electrodynamics were in fact performed using noncovariant gauges
\cite{Schwinger:1958}.

Finally, it should be noted that an impressive amount of evidence 
has been accumulated to support the consistency of this particular 
gauge. This evidence includes:
\begin{itemize}
\item{An explicit check of the tree order gravitational Ward
identity \cite{Tsamis:1992zt};}
\item{An explicit check of the one graviton loop gravitational Ward 
identity \cite{Tsamis:1996qk};}
\item{A detailed physical comparison between the noncovariant gauge
result for the one (photon) loop contribution to a charged scalar 
self-mass-squared \cite{Kahya:2005kj,Kahya:2006ui} and the analogous
covariant guage result \cite{Prokopec:2006ue}; and}
\item{A computation of the linearized Weyl-Weyl correlator in both
the noncovariant \cite{Mora:2012kr} and covariant gauges 
\cite{Mora:2012zh}.}
\end{itemize}

\section{Appendix B: Integrals from Section \ref{solve}}\label{appendix}

The purpose of this appendix is to summarize the necessary details for evaluating 
the nonlocal terms (\ref{integl1}-\ref{integl3}), and show they indeed make no 
significant contribution in the late time regime.
The first step is to perform the angular integrations using the 
formula,
\begin{equation}
\int \!\! d^3x' \Theta f(\Delta x) e^{-i\vec{k} \cdot \Delta \vec{x}}
=4\pi \!\! \int_0^{\Delta \eta} \!\!\!\!\! dr r^2 f(r)
\frac{\sin(kr)}{kr} \; . \label{angular}
\end{equation}
Employing relation (\ref{angular}) in (\ref{integl1}-\ref{integl3}) 
allows us to write,
\begin{eqnarray}
I_1 & \!\!\!\!\! =\!\!\!\!\! & -\frac{a^{-1} (\partial_0^2 \!+\!
k^2)^3}{96 \pi^2 k} \!\! \int^{\eta}_{\eta_i} \!\! d\eta' 
\frac{u(\eta',k)}{a'} \!\! \int^{\Delta \eta}_{0} \!\!\!\!\! dr r
\sin(kr) \Biggl\{\ln\! \left[\f{H^2}{4} (\Delta \eta^2 \!-\! r^2) \right]
\!-\! 1 \Biggr\} , \qquad \label{integl1sim1} \\ 
I_2 & \!\!\!\!\! = \!\!\!\!\! & -\frac{H^2 (\partial_0^2 \!+\! k^2)^2
}{32\pi^2k} \!\! \int^{\eta}_{\eta_i} \!\! d\eta' u(\eta',k) \!\!
\int^{\Delta \eta}_{0} \!\!\!\!\! dr r \sin(kr) 
\Biggl\{\ln\! \left[\f{H^2}{4} (\Delta \eta^2 \!-\! r^2) \right] 
\!-\! 1\Biggr\} , \qquad \label{integl2sim1} \\
I_3 & \!\!\!\!\! = \!\!\!\!\! & \frac{H^2(\partial_0^2 \!+\! k^2) 
\partial_0^2}{8 \pi^2 k} \!\! \int^{\eta}_{\eta_i} \!\! d\eta' u(\eta',k)
\!\! \int^{\Delta \eta}_{0} \!\!\!\!\! dr r \sin(kr) \Biggl\{\ln\!
\left[\frac{H^2}{4} (\Delta \eta^2 \!-\! r^2) \right] \!+\!1 \Biggr\}
, \label{integl3sim1} 
\end{eqnarray}
where $\eta_i \equiv -H^{-1}$ denotes the initial time. The next step 
is to perform the two independent radial integrations,
\begin{eqnarray}
J_1(\Delta \eta,k) & \!\!\!\!\! \equiv \!\!\!\!\! & \int^{\Delta \eta}_0 
\!\!\!\!\! dr r \sin(kr) = \frac{T(k \Delta \eta)}{k^2}
\label{integlJ1} \; , \\
J_2(\Delta \eta,k) & \!\!\!\!\! \equiv \!\!\!\!\!\! & 
\int^{\Delta \eta}_0 \!\!\!\!\! dr r \sin(kr) \ln\left[\f{H^2}{4}
(\Delta \eta^2 \!-\! r^2) \right] \; , \\
& \!\!\!\!\! = \!\!\!\!\! & \frac{2 \ln(H\Delta \eta)}{k^2} \, 
T(k \Delta \eta) + \frac{1}{k^2} \!\! \int^1_0 \!\! \frac{dt}{t}
\Biggl\{ T\left[k \Delta \eta (1 \!-\! 2t) \right] \!-\! 
T(k \Delta \eta) \Biggr\} \; , \qquad \label{integlJ2} 
\end{eqnarray}
where $T(x) \equiv \sin(x) - x \cos(x) = \frac13 x^3 + O(x^5)$. This
allows us to express the integrals $I_{1-3}(\eta,k)$ as,
\begin{eqnarray}
I_1(\eta,k) & \!\!\!\!\! = \!\!\!\!\! & -\frac{a^{-1}(\partial_0^2 
\!+\! k^2)^3}{96 \pi^2 k} \!\! \int^{\eta}_{\eta_i} \!\! d\eta'
\frac{u(\eta',k)}{a'} \Biggl\{ J_2(\Delta \eta,k) \!-\!
J_1(\Delta \eta,k) \Biggr\} \; , \label{integl1sim2} \\
I_2(\eta,k) & \!\!\!\!\! = \!\!\!\!\! & -\frac{H^2 (\partial_0^2 \!+\!
k^2)^2}{32 \pi^2 k} \!\! \int^{\eta}_{\eta_i} \!\! d\eta' 
u(\eta',k) \Biggl\{J_2(\Delta \eta,k) \!-\! J_1(\Delta \eta,k)
\Biggr\} \; , \label{integl2sim2} \\
I_3(\eta,k) & \!\!\!\!\! = \!\!\!\!\! & \frac{H^2 \partial_0^2
(\partial_0^2 \!+\! k^2)}{8 \pi^2 k} \!\! \int^{\eta}_{\eta_i} \!\! 
d\eta' u(\eta',k) \Biggl\{J_2(\Delta \eta,k) \!+\! 
J_1(\Delta \eta,k) \Biggr\} \; . \label{integl3sim2} 
\end{eqnarray}

Because the various integrands of (\ref{integl1sim2}-\ref{integl3sim2})
vanish like $\Delta \eta^3 \ln(\Delta \eta)$ at $\eta' = \eta$ we can 
pass one factor of the differential operator $(\del_0^2+k^2)$ through 
the integration to act on $J_1(\Delta \eta,k)$ and $J_2(\Delta \eta,k)$,
\begin{eqnarray}
\lefteqn{
I_1(\eta,k) = -\frac{a^{-1} (\partial_0^2 \!+\! k^2)^2}{48 \pi^2 k} \!\!
\int^{\eta}_{\eta_i} \!\! d\eta' \frac{u(\eta',k)}{a'} 
\Biggl\{ 2 \sin(k \Delta \eta) \ln(H \Delta \eta) } \nonumber \\
& & \hspace{5cm} + \int^1_0 \!\! \frac{dt}{t} \Bigl[\sin\left[k 
\Delta \eta (1 \!-\! 2 t) \right] \!-\! \sin(k \Delta \eta) \Bigr]
\Biggr\} , \qquad \label{integl1sim3} \\
\lefteqn{
I_2(\eta,k) = -\frac{H^2 (\partial_0^2 \!+\! k^2)}{16 \pi^2 k} \!\!
\int^{\eta}_{\eta_i} \!\! d\eta' u(\eta',k) \Biggl\{
2 \sin(k \Delta \eta) \ln(H \Delta \eta) } \nonumber \\
& & \hspace{5cm} + \int^1_0 \!\! \frac{dt}{t} \Bigl[ \sin\left[k
\Delta \eta (1 \!-\! 2 t) \right] \!-\! \sin(k \Delta \eta) \Bigr]
\Biggr\} , \qquad \label{integl2sim3} \\
\lefteqn{
I_3(\eta,k) = \frac{H^2 \partial_0^2}{4 \pi^2 k} \!\! 
\int^{\eta}_{\eta_i} \!\! d \eta' u(\eta',k)
\Biggl\{2\sin(k \Delta \eta) \ln(H \Delta \eta) + 
2 \sin(k \Delta \eta) } \nonumber \\
& & \hspace{5cm} + \int^1_0 \!\! \frac{dt}{t} \Bigl[ \sin\left[k
\Delta \eta (1 \!-\! 2 t) \right] \!-\! \sin(k \Delta \eta) \Bigr]
\Biggr\} \, . \qquad \label{integl3sim3} 
\end{eqnarray}
We can pass one more derivative through the integration using the
identities,
\begin{eqnarray}
(\partial_0^2 \!+\! k^2) \Bigl[ u(\eta',k) f(\Delta \eta)\Bigr] & = &
-(\partial_0 \!-\! i k) \times \partial_0' \Bigl[u(\eta',k) 
f(\Delta \eta)\Bigr] \; , \qquad \\
(\partial_0^2 \!+\! k^2)^2 \Bigl[ \frac{u(\eta',k)}{a'} 
f(\Delta \eta)\Bigr] & = & -(\partial_0^2 \!+\! k^2) 
(\partial_0 \!-\! i k) \times \partial_0' \Bigl[\frac{u(\eta',k)}{a'} 
f(\Delta \eta)\Bigr] \nonumber \\
& & \hspace{1cm} + H (\partial_0 \!-\! i k)^2 \times \partial_0'
\Bigl[u(\eta',k) f(\Delta \eta)\Bigr] \; . \qquad
\end{eqnarray}
The final results are (\ref{integl5ans}-\ref{integl7ans}).

\centerline{\bf Acknowledgements}

We are grateful for conversation and correspondence on this subject
with S. Deser, K. E. Leonard, S. P. Miao and T. Prokopec. This work was 
partially supported by NSF grant PHY-1205591 and by the Institute for 
Fundamental Theory at the University of Florida.

\end{document}